\documentclass[a4paper,11pt]{article}
\pdfoutput=1
\usepackage{jheppub}

\usepackage{hyperref}
\hypersetup{
     colorlinks=true,       
     linkcolor=red,          
     citecolor=magenta,        
     filecolor=magenta,      
     urlcolor=blue           
}

\usepackage{graphicx}
\usepackage{placeins}
\usepackage{url}
\usepackage{multirow}
\usepackage{paralist}
\usepackage{amssymb}
\usepackage{amsmath}
\usepackage{xspace}
\usepackage{booktabs}
\usepackage[symbol]{footmisc}

\usepackage{lineno}

\newcommand{\newc}{\newcommand}
\newc{\cpp}{\textsf{C++}}
\newc{\HWS}{\textsf{Herwig 7}}
\newc{\HW}{\textsf{Herwig}}
\newc{\HSeven}{\textsf{H7}}
\newc{\HWHML}{\textsf{H7+HADML}}
\newc{\ThePEG}{\textsf{ThePEG}}
\newc{\HADML}{\textsf{HADML}}
\newc{\HWPPClass}[1]{\mbox{\href{http://projects.hepforge.org/herwig/doxygen/classHerwig_1_1#1.html}{\textsf{#1}}}}
\newc{\NC}{N_{\mathrm{c}}}

\begin{document}
\newcommand{\xju}[1]{\textcolor{red}{#1}}
\newcommand{\bpn}[1]{\textcolor{red}{#1  --bpn}}

\title{Integrating Particle Flavor into\\Deep Learning Models for Hadronization}

\affiliation[a]{Scientific Data Division, Lawrence Berkeley National Laboratory, Berkeley, CA 94720, USA}
\affiliation[b]{Physics Division, Lawrence Berkeley National Laboratory, Berkeley, CA 94720, USA}
\affiliation[c]{Berkeley Institute for Data Science, University of California, Berkeley, CA 94720, USA}
\affiliation[d]{Department of Physics, University of California, Berkeley, CA 94720, USA}
\affiliation[e]{Jagiellonian University, Krakow, Poland}
\author[a]{Jay Chan,}
\author[a]{Xiangyang Ju,}
\author[e]{Adam Kania,}
\author[b,c]{Benjamin Nachman,}
\author[d,b]{Vishnu Sangli,}
\author[e]{and Andrzej Siodmok}

\abstract{
Hadronization models used in event generators are physics-inspired functions with many tunable parameters.  Since we do not understand hadronization from first principles, there have been multiple proposals to improve the accuracy of hadronization models by utilizing more flexible parameterizations based on neural networks.  These recent proposals have focused on the kinematic properties of hadrons, but a full model must also include particle flavor.  In this paper, we show how to build a deep learning-based hadronization model that includes both kinematic (continuous) and flavor (discrete) degrees of freedom.  Our approach is based on Generative Adversarial Networks and we show the performance within the context of the cluster hadronization model within the Herwig event generator.



}

\maketitle

\section{Introduction}
\label{sec:intro}

Despite the extensive predictability of Quantum Chromodynamics, the theory of the strong force, we cannot yet calculate how the fundamental degrees of freedom (quarks/gluons) combine to form the observable states (hadrons). At the same time, it is essential that we be able to model this transition in order to connect first-principles, perturbative calculations with data.  Event generators in wide use are based on one of two physically-inspired parametric models with many tunable parameters: the cluster model~\cite{Webber:1983if} (default in Herwig~\cite{Corcella:2000bw,Bahr:2008pv,Bellm:2015jjp,Bellm:2019zci} and Sherpa~\cite{Gleisberg:2008ta,Sherpa:2019gpd}) and the string model~\cite{Andersson:1983ia,Sjostrand:1984ic} (default in Pythia~\cite{Sjostrand:2007gs,Sjostrand:2006za}).  These models have enabled a wide array of physics results across particle and nuclear physics.  However, it is also well-known that these models do not describe all regions of phase space. As the science requires more precision and examines more extreme regions of phase space, new and systematically improvable hadronization models are needed.

Deep generative neural networks are promising tools to enhance the precision of hadronization models due to their flexibility~\cite{Butter:2022rso}.  There is a long history of neural networks for modeling non-perturbative functions~\cite{NNPDF:2021njg} and recent studies have shown that deep generative models can emulate the string and cluster hadronization when trained on paired sets of partons and hadrons~\cite{Ilten:2022jfm,Ghosh:2022zdz}.  These techniques have also been extended to the realistic setting where a pairing is not known and only hadrons are observed~\cite{Chan:2023ume,Bierlich:2023zzd}. However, all of the studies so far have trained on simplified simulations without any parton/hadron flavor.

In this paper, we extend the HadML~\cite{Ghosh:2022zdz,Chan:2023ume} setup to include parton and hadron flavor.  This is challenging because both continuous (kinematic) and discrete (flavor) information must be generated at the same time.  HadML is based on a Generative Adversarial Network (GAN)~\cite{Goodfellow:2014:GAN:2969033.2969125,Creswell2018} because it naturally accommodates the realistic case mentioned above\footnote{In fact, one can also view the MLHad approach~\cite{Bierlich:2023zzd} as a GAN where the generator is parameterized as a normalizing flow~\cite{10.5555/3045118.3045281,Kobyzev2020} and the discriminator is similar to the Wasserstein GAN setup~\cite{pmlr-v70-arjovsky17a}.  They also proposed a clever variation to avoid regenerating events in each epoch of training through reweighting.}~\cite{Chan:2023ume}.

This paper is organized as follows.  Section~\ref{sec:methods} introduces the machine learning methods of the new HadML model and how they are implemented in practice.  The dataset we use to stress-test this model, based on \textsc{Herwig}~7 (H7)~\cite{Bellm:2019zci}, is described in Sec.~\ref{sec:dataset} and numerical results are presented in Sec.~\ref{sec:results}.  The paper ends with conclusions and outlook in Sec.~\ref{sec:conclusions}.

\section{Methods}
\label{sec:methods}

\subsection{Generative Adversarial Network Framework}

The overall setup is similar to that in the previous work \cite{Ghosh:2022zdz}. A conditional generator function $G\left(z, \lambda; \omega_G \right)$ with the parameters $\omega_G$ is learned to map the initial cluster properties onto the properties of the two\footnote{The decay of heavy clusters can produce more than two hadrons, but in most cases the collisions we consider produce mainly light clusters that decay into two hadrons. Therefore, in this study, we have limited ourselves to the case of decay into two hadrons, and we will investigate more complex decays in future work.} hadrons from each cluster decay $\{h_1, h_2\} \in\mathbb{R}^{2 N_h}$. In addition to the four-momenta, the generator function should also output the particle types of the two hadrons. Here, $z\in\mathbb{R}^{N_z}$ is the input noise variable sampled from the prior $p\left(z\right)$, and $\lambda\in\mathbb{R}^{N_\lambda}$ is the conditional variable. In Ref~\cite{Ghosh:2022zdz}, the generator function was conditioned on the cluster four-momentum ({$E$, $p_x$, $p_y$, $p_z$}). In this paper, we consider additional conditional variables from the two incoming cluster-forming quarks, including their four-momenta and particle types. Since two hadrons from a cluster decay must be back-to-back in the rest frame of cluster, the generator $G$ can output the polar angles $\theta$ and $\phi$ of the ``first hadron'' in the cluster rest frame instead of the 4-momenta of both hadrons. Similarly, the incoming quarks are back-to-back in the cluster rest frame. Therefore, we parametrize their four-momentum as the polar angles $\theta$ and $\phi$ of the ``first quark'' in the cluster rest frame. Note that here $\phi$ is defined in the range of $(-\pi/2,\ \pi/2)$, and the hadron (quark) with $\phi$ in this range is defined to be the first hadron (quark).

A discriminator function $D\left(\theta_{h_1}, \phi_{h_1}, \mathrm{PID}_{h_1}, \mathrm{PID}_{h_2}; \omega_D \right)$, parameterized with $\omega_D$, is learned to represent the probability that $\{\theta_{h_1}, \phi_{h_1}, \mathrm{PID}_{h_1}, \mathrm{PID}_{h_2}\}$ came from cluster fragmentation rather than the generator $G$. Note that $\theta_{h_1}$, $\phi_{h_1}$ are the polar angles of the first hadron, and $\mathrm{PID}_{h_1}$, $\mathrm{PID}_{h_2}$ are the particle types of the two hadrons. $G$ and $D$ are trained alternately where $G$ is trained to maximize the loss function:

\begin{equation}
    L_G = - \mathbb{E}_{\lambda \sim \text{H7},\, z \sim p\left(z\right)} \big( \log\left(D\left(\tau\left(\lambda\right)\right)\right) + \log\left(1 - D\left(G\left(z, \lambda\right)\right)\right) \big)\,,
\end{equation}
where $\tau$ is the cluster fragmentation, and D is trained to minimize the loss function:

\begin{equation}
    L_D = - \mathbb{E}_{\lambda \sim \text{H7},\, z \sim p\left(z\right)} \big( \log\left(D\left(\tau\left(\lambda\right)\right)\right) + \log\left(1 - D\left(G\left(z, \lambda\right)\right)\right) \big) + \gamma R_1(\omega_D) \,,
\end{equation}
where $\gamma$ is a regularization weight, which we set to 200. We use R1 regularization \cite{GANconverge} on real data points:
\begin{equation}
R_1(\omega_D) = \mathbb{E}_{\lambda \sim \text{H7}}[\| \nabla_{\tau(\lambda)} D_{\omega_D} (\tau(\lambda)) \|^2].
\end{equation}

\subsection{Machine Learning Implementation}

The generator ($G$) and discriminator ($D$) functions are both parametrized as neural networks. Each of them is a fully connected network with four hidden layers, each with a width of 1,000 neurons. All intermediate layers in these networks use a LeakyReLU~\cite{xu2015empirical} activation function.

The non-discrete conditional inputs of $G$ are normalized to the range of $(-1,\ 1)$, whereas the noise prior $p$ is a Gaussian distribution with a mean of 0 and width of 1. The noise dimension $N_z$ is set to 64. The last layer of $G$ is divided into the variables $\kappa \in\mathbb{R}^{2}$, which correspond to hadron kinematics, and the variables $\pi \in\mathbb{R}^{2 N_t}$, which correspond to hadron types. Here, $N_t$ is the number of hadron types considered. For simplicity, we consider only the 40 most common hadron types (i.e. $N_t = 40$)\footnote{The most common hadron types are identified from an independent and slightly different simulation sample generated by H7.  This is why the frequencies reported in Fig.~\ref{fig:pdgid1} are not strictly decreasing.}. The hadron kinematics $\theta_{h_1}$ and $\phi_{h_1}$ are extracted from $\kappa$ with a tanh activation function, as in the previous work \cite{Ghosh:2022zdz}. The hadron type, on the other hand, is a categorical variable. In order to avoid zero gradients when using {\it argmax} in training, we use the Gumbel-Softmax \cite{jang2017categorical} distribution to approximate the distribution of hadron types:
\begin{equation}
    y_i = \frac{\exp\left(\left(\log\pi_i + g_i\right) / \tau\right)} {\sum_i \exp\left(\left(\log\pi_i + g_i\right) / \tau\right)},
\end{equation}
where $g_i$ are independent and identically distributed samples drawn from Gumbel(0, 1). $\tau$ is a temperature parameter and as it approaches 0 the Gumbel-Softmax distribution becomes identical to the categorical distribution. We anneal $\tau$ by linearly decreasing it from 1.0 to 0.1 during training. The hadron type distributions $y$ from the Gumbel-Softmax distribution are then taken as the inputs for $D$, in addition to the hadron kinematics $\theta_{h_1}$ and $\phi_{h_1}$. During inference, the generated hadron types are obtained from the Gumbel-Softmax distribution with the {\it argmax} operation. The last layer of $D$ uses a sigmoid activation function.

All neural networks are implemented and trained using PyTorch \cite{NEURIPS2019_9015}. The generator and discriminator are optimized alternately with Adam \cite{adam} with a learning rate of $3 \times 10^{-4}$ for both networks. The training uses a batch size of 40,000 and is performed for 25 epochs. The hyperparameters are optimized with Weights and Biases~\cite{wandb}.

\section{Dataset}
\label{sec:dataset}
The dataset was generated by the \textsc{Herwig} 7.2.1 Monte Carlo generator, which by default uses a cluster hadronization model. In the first step of the cluster model, partons are grouped into colorless objects called clusters (exited pre-hadrons), which then decay into two hadrons (or lighter clusters\footnote{The heavier clusters can also decay into lighter clusters before decaying into hadrons. However, since in this publication, we are mainly interested in the generation of individual hadron flavour. We leave the decays of the heavy clusters for a future follow-up paper.}).
Since in our study, we wanted to integrate the particle flavour into the HadML model, in addition to the kinematic information (the four momenta of all light clusters in the event and their decay products, hadrons), our dataset also contains information about the type of hadrons ($\mathrm{PID}_{h_1}$, $\mathrm{PID}_{h_2}$), as well as the Particle Data Group~\cite{ParticleDataGroup:2022pth} Identification (PDG ID) of the partons that make up a given light cluster. All datasets were generated in electron-positron collisions at an energy of 91.2 GeV, which corresponds to events recorded by LEP experiments at CERN. Data from LEP is crucial for fitting hadronisation models, therefore such a sample is the most natural for the development of new hadronization approaches.
To test whether the HadML model can adapt to different flavor compositions, we prepared two datasets with different settings of the cluster model parameters responsible for the generation of hadron types.
To be more precise, the nominal dataset was generated using H7's default settings. For the variational dataset, we have maximized the weights for producing charmed quark-antiquark pairs, strange quark-antiquark pairs and diquark-antiquark pairs as well as the relative weight $SngWt$ for the production of singlet baryons and the relative weight $DecWt$ for the production of decuplet baryons in cluster hadronisation\footnote{
To achieve this, we used the following H7's settings: 
HadronSelector:PwtDIquark=10, HadronSelector:PwtBquark=10, HadronSelector:PwtCquark=10, 
HadronSelector:SngWt=10 and 
HadronSelector:DecWt=10. For details, please see H7's manual~\cite{Bahr:2008pv}.}.

\section{Results}
\label{sec:results}

\begin{figure}
    \centering
    \includegraphics[width=0.7\textwidth]{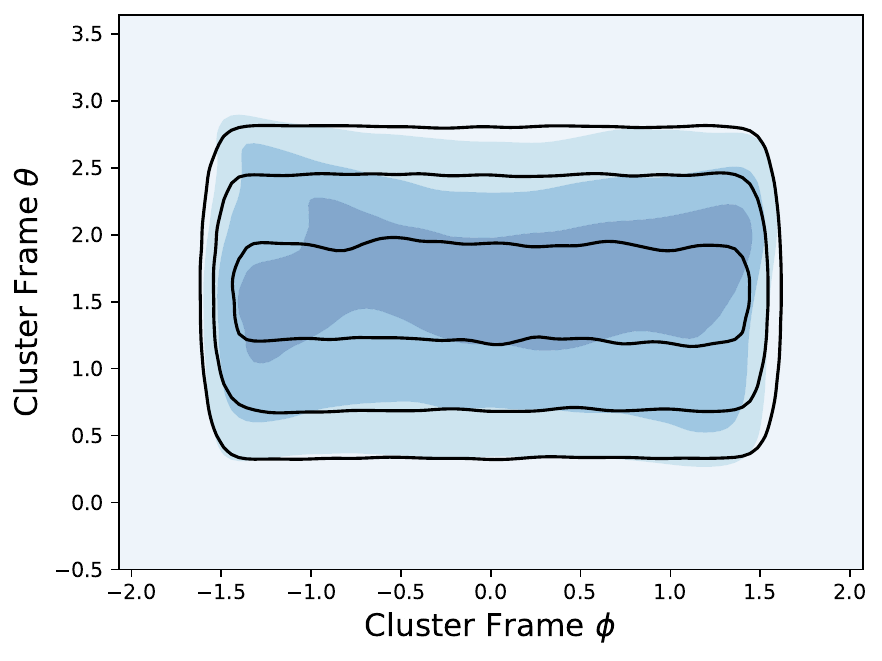}
    \caption{Flavor-inclusive density contour of $\phi$ and $\theta$ in the cluster frame for the nominal dataset (solid line) as well as for the dataset created from the fitted HadML model (filled colors).}
    \label{fig:kinematics}
\end{figure}

The two datasets described in Sec.~\ref{sec:dataset} have the same distributions of $\theta_h$ and $\phi_h$, which are nearly independent of hadron type.  As in our previous works, these distributions are well-modeled by HadML (Fig.~\ref{fig:kinematics}) and are essentially uniform in $\phi$ and Gaussian-like in $\theta$.

\begin{figure}
    \centering
    \includegraphics[width=0.7\textwidth]{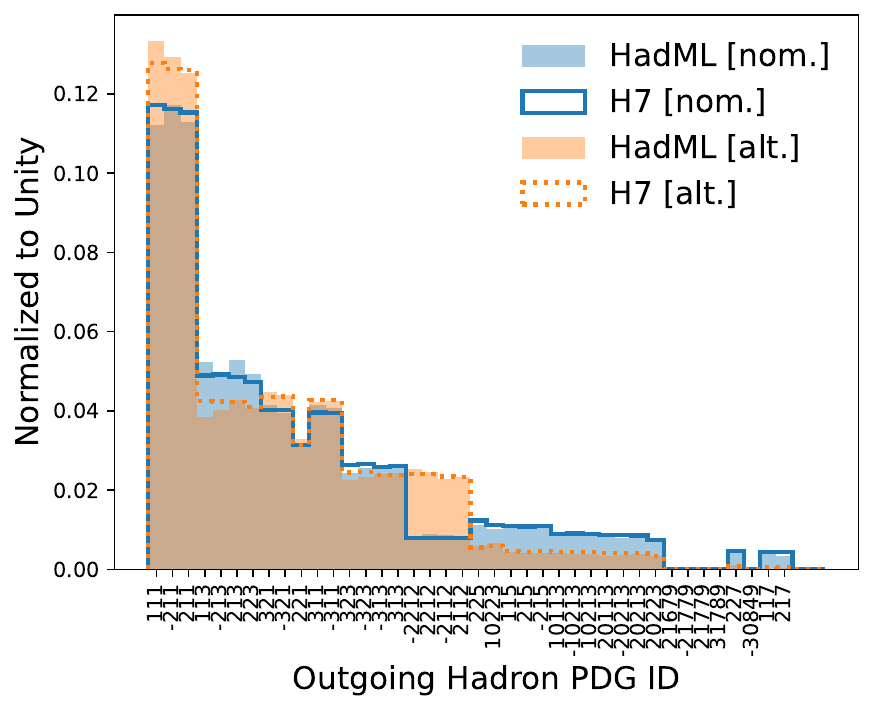}
    \caption{The Particle Data Group Identification (PDG ID)~\cite{ParticleDataGroup:2022pth} of hadrons generated by the nominal and alternative H7 datasets as well as for datasets created from the fitted HadML models.}
    \label{fig:pdgid1}
\end{figure}

The lab-frame spectra of energies and angles differ between the nominal and alternative samples because of the differences in hadron masses.  Since the masses are known, the lab-frame properties are therefore determined by how well we model the frequency of the various flavor types.  Figure~\ref{fig:pdgid1} shows the frequencies for H7 and HadML, inclusive in the flavor of the quark types composing the decaying cluster.  As expected, the pions are the most frequent (PID IDs 111 and $\pm211$), followed by the $\rho$ (PDG IDs 113 and $\pm213$) and $\omega$ (PDG ID 223).  Next are the kaons (PDG ID 3xx), protons (PDG ID $\pm 2212$), and neutrons (PDG ID $\pm 2112$).  A series of other intermediate hadronic resonances follow the lightest baryons.  The nominal and alternative H7 models significantly differ in these rates, most notably for the pion versus $\omega$ production and in the rate of baryons.  HadML captures these trends across the full spectrum at $\mathcal{O}(1\%)$ precision.  This is true even for the large drop in frequency between the pions and all other hadrons as well as between the large raise in baryon production between the nominal and alternative H7 models.

\begin{figure}
    \centering
    \includegraphics[width=0.5\textwidth]{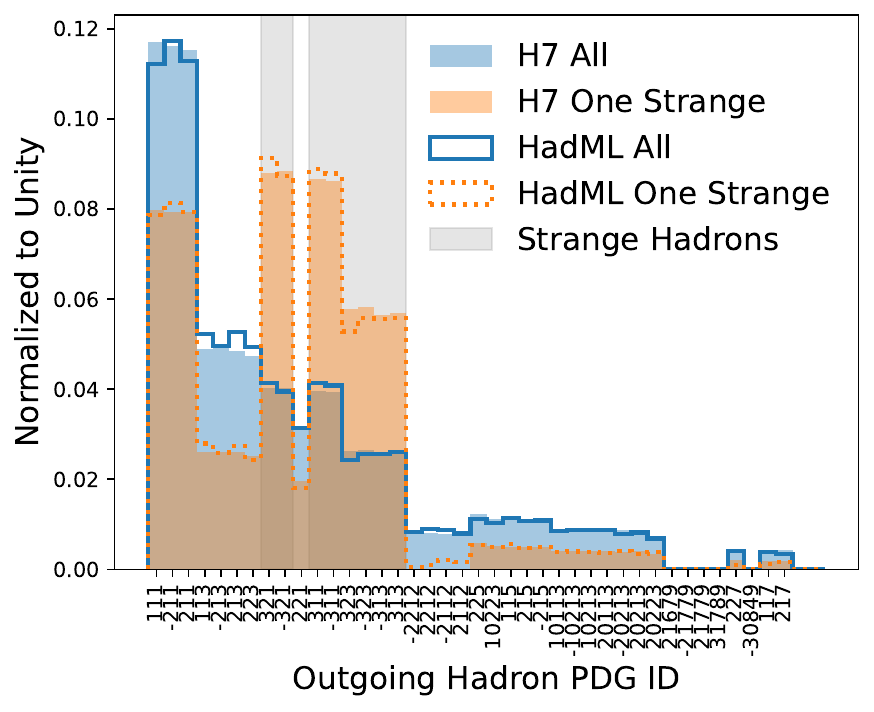}\includegraphics[width=0.5\textwidth]{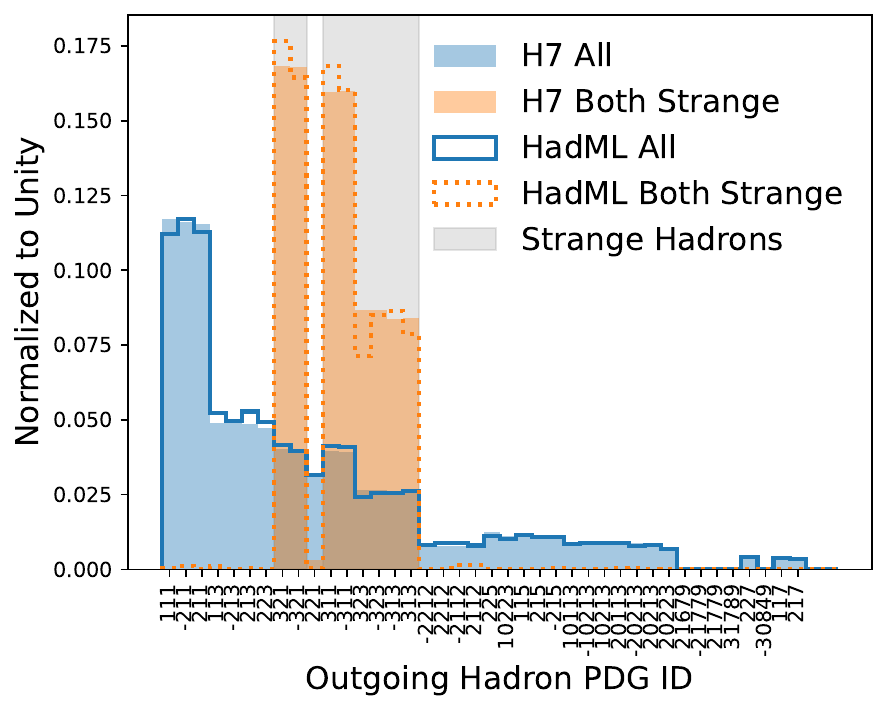}
    \caption{Left (Right): The PDG ID of hadrons generated inclusively and requiring that at least one (both) of the quarks composing the cluster are strange for H7 and for the HadML model. }
    \label{fig:pdgid2}
\end{figure}

Since HadML is conditioned on the flavor of the constituent quarks, we can also investigate the relationship between the cluster composition and the resulting hadron types.  We expect that if one of the quarks is strange, then the hadrons should be biased towards strange hadrons (e.g. kaons).  This is what we see in Fig.~\ref{fig:pdgid2}, which compares the inclusive hadron flavor distribution with the spectrum after requiring that at least one of the incoming quarks is strange.  There are large changes between the inclusive and conditional distributions, which are well-reproduced by HadML.  The neural network is also able to learn that when both quarks are strange, then both outgoing hadrons must be strange.

\section{Conclusions}
\label{sec:conclusions}

This paper marks an important milestone in the development of surrogate models for hadronization: the inclusion of hadron flavor.  Hadronization is not understood from first principles so it is a natural candidate for the flexible modeling afforded by deep generative models.  Previous works on machine learning-based hadronization had focused on the case of only pions and we extend the GAN-based HadML approach to include other hadron types\footnote{While this work was being finalized, Ref.~\cite{Birk:2023efj} became the first paper to combine continuous and discrete outputs for generative modeling in HEP.  They used a diffusion model with only a few discrete labels (particle types), both of which differ from (and are not directly applicable to) our setup.  However, it would be interesting to explore possible connections between approaches in the future.}.  This required us to combine continuous (e.g. angles/momenta) with discrete (particle types) in our generator.  We accomplished this goal using the Gumbel-Softmax~\cite{jang2017categorical} distribution for hadron types, to enable differentiability.  

The insights of this paper could be combined with our previous paper~\cite{Chan:2023ume} to fit the HadML model with flavor to data in the lab frame.  Additional work is also required to integrate all hadron types and to go beyond two-body decays of hadrons.  Ultimately, we hope to create a model flexible enough to accommodate the cluster model, the string model, and nature.

\section*{Software and Datasets}

The nominal and alternative H7 samples used for training can be found on Zenodo at \href{https://zenodo.org/records/10246934}{https://zenodo.org/records/10246934} \cite{chan_2023_10246934}.  Software for reproducing the plots can be found on Github at \href{https://github.com/hep-lbdl/hadml/releases/tag/2.0.0}{https://github.com/hep-lbdl/hadml/releases/tag/2.0.0} \cite{chan_2023_10275487}.

\section*{Acknowledgments}
We thank Aishik Ghosh for many useful discussions. The work of AS is funded by grant no. 2019/34/E/ST2/00457 of the National Science Centre,
Poland. A.K. acknowledges support by the Priority Research Area Digiworld under the program Excellence Initiative
– Research University at the Jagiellonian University in Cracow.
JC, BN and XJ are supported by the U.S.\ Department of Energy (DOE), Office of Science under contract number DE-AC02-05CH11231. Support for JC and XJ was also provided through the Scientific Discovery through Advanced Computing (SciDAC) program funded by U.S. Department of Energy, Office of
Science, Advanced Scientific Computing Research and High Energy Physics.




\FloatBarrier
\bibliographystyle{JHEP}
 \bibliography{HEPML,other-refs}

\end{document}